\documentclass[runningheads,a4paper]{llncs}
\usepackage[utf8]{inputenc}
\usepackage[T1]{fontenc}
\usepackage{graphicx}
\usepackage{hyperref}
\usepackage{url}
\usepackage{tipa}
\usepackage[english]{babel}

\DeclareRobustCommand{\openE}{\textipa{E}}
\DeclareRobustCommand{\implB}{\texthtb} 
\DeclareRobustCommand{\palN}{\textltailn}
\DeclareRobustCommand{\openO}{\textopeno} 

\title{Generative Artificial Intelligence, Musical Heritage and the Construction of Peace Narratives: A Case Study in Mali}
\titlerunning{GenAI, Musical Heritage and Peace in Mali}
\author{Nouhoum Coulibaly\inst{1} \and Ousmane Ly\inst{2} \and Michael Leventhal\inst{1} \and Ousmane Goro\inst{3}}
\authorrunning{N Coulibaly, O Ly, M Leventhal, O Goro}
\institute{RobotsMali AI4D Lab, Bamako, Mali \\
\email{research@robotsmali.org}\and Faculté de Médecine et d’Odonto-Stomatologie (FMOS), Université des Sciences, Techniques et Technologies de Bamako (USTTB), Bamako, Mali \and Conservatoire des Arts et Métiers Multimédia Balla Fasseké, Bamako, Mali}

\begin{document}
\maketitle
\begin{abstract}
This study explores the capacity of generative artificial intelligence (Gen AI) to contribute to the construction of peace narratives and the revitalization of musical heritage in Mali. The study has been made in a political and social context where inter-community tensions and social fractures motivate a search for new symbolic frameworks for reconciliation. The study empirically explores three questions: (1) how Gen AI can be used as a tool for musical creation rooted in national languages and traditions; (2) to what extent Gen AI systems enable a balanced hybridization between technological innovation and cultural authenticity; and (3) how AI-assisted musical co-creation can strengthen social cohesion and cultural sovereignty. The experimental results suggest that Gen AI, embedded in a culturally conscious participatory framework, can act as a catalyst for symbolic diplomacy, amplifying local voices instead of standardizing them. However, challenges persist regarding the availability of linguistic corpora, algorithmic censorship, and the ethics of generating compositions derived from copyrighted sources.
\end{abstract}
\section{Introduction: Context and Problem Statement}
Mali faces persistent challenges in terms of social cohesion. A country once renowned for its inter-ethnic respect and tolerance, now confronts an increasingly fragmented cultural identity threatening the tradition of peaceful co-existence. There is a national imperative to discover narrative innovations capable of recomposing the societal fabric. Music, an important, historic medium for transmitting identity and cultural mediation, sometimes appears insufficient to reach modern audiences or engage with the aspirations of younger generations when it remains confined to traditional forms.

The emergence of generative artificial intelligence (Gen AI) opens new possibilities for exploring the nexus of cultural heritage and innovation to produce a modern language of reconciliation imbued with local meaning. Suno AI, a platform for generating musical compositions from text prompts~\cite{CasiniEtAl2025}, was selected as a relevant tool for exploring this intersection of technology and tradition. The complementary use of ChatGPT and GEMINI for the generation and translation~\cite{CasiniEtAl2025} of texts into national languages completed the technological framework for the experiment.

This study tested the hypothesis that Gen AI, used responsibly with cultural sensitivity~\cite{pramMorreale2025}, can revitalize musical heritage by fostering the emergence of indigenous, socially inclusive, and technologically contemporary peace narratives.

\section{Methodology}
\subsection{Study design and Research Questions}

The study employed an empirical qualitative approach centered on a co-creation workshop, complemented by thematic analyses of the resulting musical productions and semi-structured interviews with participants. This approach allows us to understand creative processes, cultural perceptions, and the potential of the approach to simulate the creation of culturally-rooted peace narratives. The study was guided by three primary hypotheses: (H1) Gen AI tools can be effectively appropriated by participants with varying levels of technological literacy when provided with structured training; (H2) AI-generated musical compositions incorporating traditional Malian instruments and national languages will be perceived as culturally authentic by participants; (H3) The collaborative process of AI-assisted music creation can foster intergenerational and interinstitutional dialogue around themes of reconciliation.

\subsection{Study Population}

The ``Voices of Reconciliation'' workshop brought together 30 participants recruited using a stratified sample ensuring diversity: 10 ministerial representatives, 9 youth actors from the National Youth Council, 6 women’s organizations, and 5 traditional music students from the Balla Fasséké Kouyaté Conservatory. This composition produced a plurality of institutional, generational, artistic and gender perspectives, essential for a credible co-construction of narratives.  Participants were required to: (a) be at least 18 years of age; (b) have basic literacy in at least one national language; (c) demonstrate interest in either traditional Malian music or digital technologies; and (d) be available for the full three-day workshop duration. Exclusion criteria included inability to commit to the complete workshop schedule.

\subsection{Workshop Structure}

The workshop was structured in three phases:

\textbf{Phase 1\ ---\ Initial training:} familiarization with the principles of Gen AI applied to music, operation of ChatGPT, GEMINI and Suno AI, introduction to prompt engineering.

\textbf{Phase 2\ ---\ Integration of tradition and modernity:} development of prompts integrating characteristic traditional Malian sound motifs (kora, balafon, djembe)~\cite{goro2025}, into contemporary arrangements, experimentation with hybrid styles.

\textbf{Phase 3\ ---\ Linguistic authenticity:} creation of lyrics in national languages (Bambara, Fulfulde, Tamasheq, Songhai, Dogon), phonetic transcription and adaptation of prompts to respect tonal structures and local cultural references.

\subsection{Creative Process and Prompt Engineering}

The creative process followed an iterative methodology combining human artistic direction with AI generation capabilities. Each composition underwent multiple refinement cycles before achieving its final form.

\textbf{Prompt Development Strategy:} Participants were trained to construct prompts using a structured framework: [Genre/Style] + [Instrumentation] + [Tempo/Mood] + [Cultural References] + [Thematic Content]. Initial prompts were often generic (e.g., \textit{African music with drums}), which yielded stereotypical outputs lacking Malian specificity.

\textbf{Successful Prompt Examples:} Through iterative refinement, participants developed more sophisticated prompts. For example, a successful prompt for the Afrobeat-Mandingo pattern was: \textit{Afrobeat fusion with traditional Mande instruments, featuring kora melody over balafon ostinato, 125 BPM, brass section with Mandingue-style horn arrangements, djembe percussion, uplifting and celebratory mood}. This specificity produced outputs that participants recognized as culturally grounded.

\textbf{Unsuccessful Approaches:} Several prompt strategies proved ineffective: (a) requests for specific artist imitations were blocked by platform content policies; (b) prompts in national languages without phonetic guidance resulted in mispronunciations; (c) overly complex prompts combining more than three musical traditions produced incoherent outputs; (d) requests for the imzad (Tuareg instrument) often defaulted to generic violin sounds due to limited training data.

\textbf{Iteration Statistics:} On average, each of the 12 final compositions required 8-12 generation attempts before achieving participant satisfaction. The primary reasons for rejection included: inappropriate instrumentation (34\%), tempo mismatches (22\%), vocal quality issues (28\%), and stylistic incongruence (16\%).

\subsection{Data collection and analysis}

The data includes twelve original compositions produced in the workshop, analyzed thematically around the concepts of peace, forgiveness, memory, diversity, and cultural mediation. The semi-structured interviews explored perceptions of legitimacy, impact, limitations, and sustainability of the approach. The entire output of the workshop is available in this \href{https://drive.google.com/drive/folders/1RLfxoWmKdjws-g_JB7muJ3w3OSDnzF4y?usp=drive_link}{Google Drive folder}.

\textbf{Data Collection Instruments:} Three complementary instruments were employed: (1) a post-workshop survey with 15 Likert-scale items and 5 open-ended questions administered to all willing participants (n=13); (2) semi-structured interviews lasting 20-30 minutes with a purposive sample of 8 participants representing each stakeholder group; and (3) systematic documentation of the creative process including all prompts attempted, generation outputs, and participant deliberations during selection.

\textbf{Analysis Framework:} Qualitative data were analyzed using thematic analysis following Braun and Clarke's six-phase framework~\cite{BraunClarke2006}. Two researchers independently coded interview transcripts, with inter-rater reliability of 0.82 (Cohen's kappa~\cite{Cohen1960}). Musical compositions were analyzed using an adapted version of Agawu's African music analysis framework~\cite{Agawu2016}, examining: rhythmic structures, melodic organization, instrumental choices, linguistic content, and symbolic references.

\section{Results}
\subsection{Hybrid musical architectures}

The analysis reveals three recurring hybridization patterns:

\textbf{Afrobeat-Mandingo pattern:} 120-130 BPM rhythmic structure in 4/4 time signature, incorporating the kora and balafon as the main melodic carriers~\cite{rouget1965}, while electric guitar, bass and brass and wind instruments (tenor saxophone, trumpet) provide the Afrobeat1 groove. Traditional passages (solo kora/balafon) alternate with energetic sections carried by electric instruments, creating a sophisticated progressive dynamic. This pattern is exemplified by \href{https://drive.google.com/file/d/1L3k-W7-y6FJytvLlq0os0DK39npCuGG5/view?usp=sharing}{this composition.} This song speaks to the solidarity between the peoples of the Sahel, making reference to the AES, the Alliance of Sahel States, a political union created in 2023 by Mali, Niger and Burkina Faso to strengthen cooperation. The French lyrics are below:

\begin{verse}
Un seul peuple, un seul cœur, AES \\
Avançons sans peur, AES \\
Dans l'unité, la paix va briller \\
Cohésion sociale on va gagner ! \\

\medskip

Jeunes et vieux marchons côte à côte \\
Nos rêves ensemble, notre force éclate \\
Chaque main qui s'élève construit demain \\
Dans l'unité, on tient notre chemin \\

\medskip

One people, one heart, AES \\
Let us move forward without fear, AES \\
In unity, peace will shine \\
Social cohesion we will win! \\

\medskip

Young and old, let us walk side by side \\
Our dreams together, our strength bursts forth \\
Each hand that rises builds tomorrow \\
In unity, we keep moving forward \\
\end{verse}

\textbf{Reggae-sabar model:} bipartite architecture featuring an initial reggae section (soft rhythm guitar, deep bass, light percussion) creating a meditative atmosphere, followed by a modulation towards Senegalese sabar rhythms characterized by energetic percussion (djembe, balafon, groove drums). This musical narrative structure symbolizes a journey from individual reflection to collective celebration. Through linguistic diversity, \href{https://drive.google.com/file/d/15z15DZ57dc2F1y4fyV5wW5ePai1V4pdf/view?usp=sharing}{this multi-language passage calls on people to join hands in collective celebration.}

\textbf{Communal harmony:} communal harmony across languages as exemplified by \href{https://drive.google.com/file/d/1THRIcgoRrvNVfVSu1QNwS2nttrFe973z/view?usp=drive_link}{this composition weaving together 5 languages.}: 

\begin{verse}
K'an b\openE{} se, k'an b\openE{} ka hakili don (Bambara) \\
Jam tun, jam tun, no \implB{}e ngoodi (Fulfulde) \\
Boro foori nda siini, bara koyra béeri (Songhay) \\
Assalam, assalam, akal negh ad tew (Tamasheq) \\
Main dans la main, bâtissons l’avenir (French) \\

\medskip

Let us be united, let us share wisdom \\
Peace everywhere, may peace guide us \\
Reconciliation and peace build the country \\
Peace, peace, may our land live \\
Hand in hand, let us build the future \\
\end{verse}

\textbf{Meditative Tuareg model:} slow tempo (80-110 BPM), Tuareg pentatonic scales, dominance of the imzad (monocord bowed string instrument) and the tende (single skin hand drum central to Tuareg culture). This model evokes a spiritual and contemplative mood~\cite{schaeffner2006}, with progressive mystical rises embodying the nomadic traditions of the desert. The Tamasheq and French lyrics present an invocation to peace for all the people of Mali. \href{https://drive.google.com/file/d/1kCwHqO0pLWXvRlLglA4yqFeDfKZYkkxj/view?usp=sharing}{Here is the composition} and Tamasheq lyrics.

\begin{verse}
Alâfiyet n Sahel \\
Alâfiyet, alâfiyet \\
Sahel wa n alâfiyet \\
Kel Tamasheq d wassaghmar \\
Amassa n tafrawt \\
Taggalt n Mali \\
Alâfiyet! Alâfiyet! \\
Sahel wa n alâfiyet \\
Taggalt d wassagher \\
Mali d Sahel \\
Alâfiyet ar abadan \\

\medskip

Peace of the Sahel \\
Peace, peace \\
Sahel of peace \\
Tuaregs and their brothers \\
Children of freedom \\
Unity of Mali \\
Peace! Peace! \\
Sahel of peace \\
Unity and brotherhood \\
Mali and Sahel \\
Peace for eternity \\
\end{verse}

\subsection{Linguistic and semantic strategies}

Four major discursive strategies structure the compositions.

\textbf{Stratified multilingualism:} the systematic use of several national languages within a single work, forging a sense of collective identity. For example, ``\textit{Y\openE{}r\openE{} ka s\openE{}b\openE{}}'' (Our Hearts are One) intersperses French with the primary language, Bambara, while ``\textit{Farafina Kelen Yan}'' (One Africa) alternates between Bambara, Fulfulde, Songhay, and Tamasheq.

\textbf{Reconciliation lexicon:} recurring lexical field structuring the compositions around key terms: \textit{b\openE{}n} (Bambara: union), \textit{fara\palN{}\openO{}kan} (Bambara: together), \textit{alâfiyet} (Tamasheq: peace), \textit{kelenya} (Bambara: unity), \textit{wassagher} (Tamasheq: brotherhood), \textit{taggalt} (Tamasheq: union), \textit{duddal} (Fulfulde: reconciliation), \textit{jam} (Fulfulde: peace). These terms carry a symbolic charge beyond their simple literal translation.

\textbf{Unifying geographical imagery:} encompassing geographical references symbolically reconstituting the territorial integrity and national identity of Mali (``From Kayes to Gao, from Ségou to Timbuktu'', ``Kidal to Ménaka'', ``From the Sahel to Bamako''), countering narratives of fragmentation. \href{https://drive.google.com/file/d/1JGNWWSNQIoM5xhy0JhViKSYER5rcyQMa/view?usp=sharing}{Here is a composition, with French lyrics, illustrating this discursive strategy:}

\begin{verse}
Je suis le Mali, Tu es le Mali, Nous sommes le Mali \\
Ensemble, nous ferons le Mali \\
Je suis le Mali, Tu es le Mali, Nous sommes le Mali \\
Debout pour le Mali \\

\medskip

Notre combat, c’est le Mali \\
Ensemble, pour toujours, le Mali! \\
De Tombouctou à Sikasso \\
Une voix qui s’élève \\

\medskip

Pour la paix, pour la terre \\
Pour l’avenir qu’on rêve \\
Nos mains unies, nos cœurs sans peur \\
Pour le Mali, pour l’honneur … \\

\medskip

I am Mali, You are Mali, We are Mali \\
Together, we will build  \\
I am Mali, You are Mali, We are Mali \\
Standing proud for Mali \\

\medskip

Our struggle is for Mali \\
Together, forever — Mali! \\
From Timbuktu to Sikasso \\
One voice rises up \\

\medskip

For peace, for our land \\
For the future of our dreams \\
Our hands united, our hearts without fear \\
For Mali, for our honor … \\
\end{verse}

\textbf{Sahelian organic metaphors:} images taken from the local environment such as the (Niger) river, \textit{ténéré n alâfiyet} (Tamasheq: desert of peace), \textit{azalay n alâfiyet} (Tamasheq: caravan of peace), and the palaver tree, anchoring the discourse of peace in daily experience and the Malian collective imagination.

\subsection{Narrative structures}

The compositions present a common narrative architecture in four phases:

\textbf{Invocation phase:} instrumental introductions and first verses establishing a historical or spiritual context, invoking ancestors, traditions or sacred geography.

\textbf{Diagnostic phase:} central verses identifying challenges without explicitly naming them, evoking ``what has been broken'' or calling for overcoming fears.

\textbf{Collective resolution phase:} refrains sung in chorus proposing unity, reconciliation, fraternity, with a repetitive character aimed at anchoring memory.

\textbf{Spiritual Elevation Phase:} musical bridges and final sections adopting a spiritual or transcendent register, with invocations such as \textit{Ya Allah! Alâfiyet!} (Tamasheq: O Allah! Peace!) or projections towards the future such as \textit{Alâfiyet ar abadan} (Tamasheq: Peace Forever) as can be \href{https://drive.google.com/file/d/1_fXpuIWmVT3RtnnWvd-mrpQqFkl3ezaG/view?usp=sharing&t=11}{heard in this song:}

\subsection{Instrumentation and cultural significance}

The organological analysis reveals a symbolic hierarchy:

\textbf{Identity instruments: } kora, balafon and djembe appear in 90\% of the compositions, functioning as markers of Mande authenticity. with the dominance of the balafon and the djembe~\cite{schaeffner2006,vanBeek2012}. \href{https://drive.google.com/file/d/1xrRwmpIrFteK8Z3mBZJ4qxnWdIYWv20p/view?usp=sharing}{Here is an illustration.}

\textbf{Mediation instruments:} brass and wind instruments and electric guitar create sounds familiar to young urbanites~\cite{hannecart2013} while respecting traditional rhythmic structures. \href{https://drive.google.com/file/d/1_Zt-zbUV0RCWarJ1igjwHyu6oSCeLGv-/view?usp=sharing}{Here is an example, ``\textit{Aw bismillah}''.}

\textbf{Specific regional instruments:} imzad and tahardent (Tuareg lute) in Tamasheq compositions, tam-tam and sabar percussion in Senegalese-style variants. \href{https://drive.google.com/file/d/1ofV4CJofgaXlVvWKqyrS2ik1SXJ5lfth/view?usp=sharing}{Here is an example.}

\subsection{Vocal performance}

The compositions systematically present a responsorial and antiphonal structure inherited from griot traditions~\cite{Agawu2016}, fulfilling several functions: participatory (choral format facilitating collective appropriation), pedagogical (repetition reinforcing memorization), and symbolic (polyphony materializing harmony in diversity)~\cite{Cousin2014CahiersNotice}.

The analysis of vocal timbres reveals the use of distinct registers for specific effects: deep male voices for solemn invocations, female voices for passages of hope and spiritual elevation, and childlike voices in certain refrains symbolizing the future.

\subsection{Technological appropriation}

Examination of the prompts used as the workshop progressed revealed a growing sophistication, allowing the specification of traditional musical scales, control of instrumental balance, integration of prosodic indications~\cite{migliore2016} respecting linguistic tones, and programming of fluid stylistic transitions. This progression testifies to a process of appropriation~\cite{hannecart2013} where the AI tool becomes an instrument at the service of a controlled cultural vision.

\subsection{Social and perceptual effects}

Participants reported a strengthening of cultural belonging  a renewed confidence in the possibilities of dialogue, and a heightened awareness of languages and traditions as positive resources. The workshop fostered institutional recognition: the works were promoted during the Semaine Nationale de la Réconciliation (National Week of Reconciliation - SENARE 2025), confirming their institutional and social resonance.

\begin{figure}[h]
  \centering
  \includegraphics[width=0.8\textwidth]{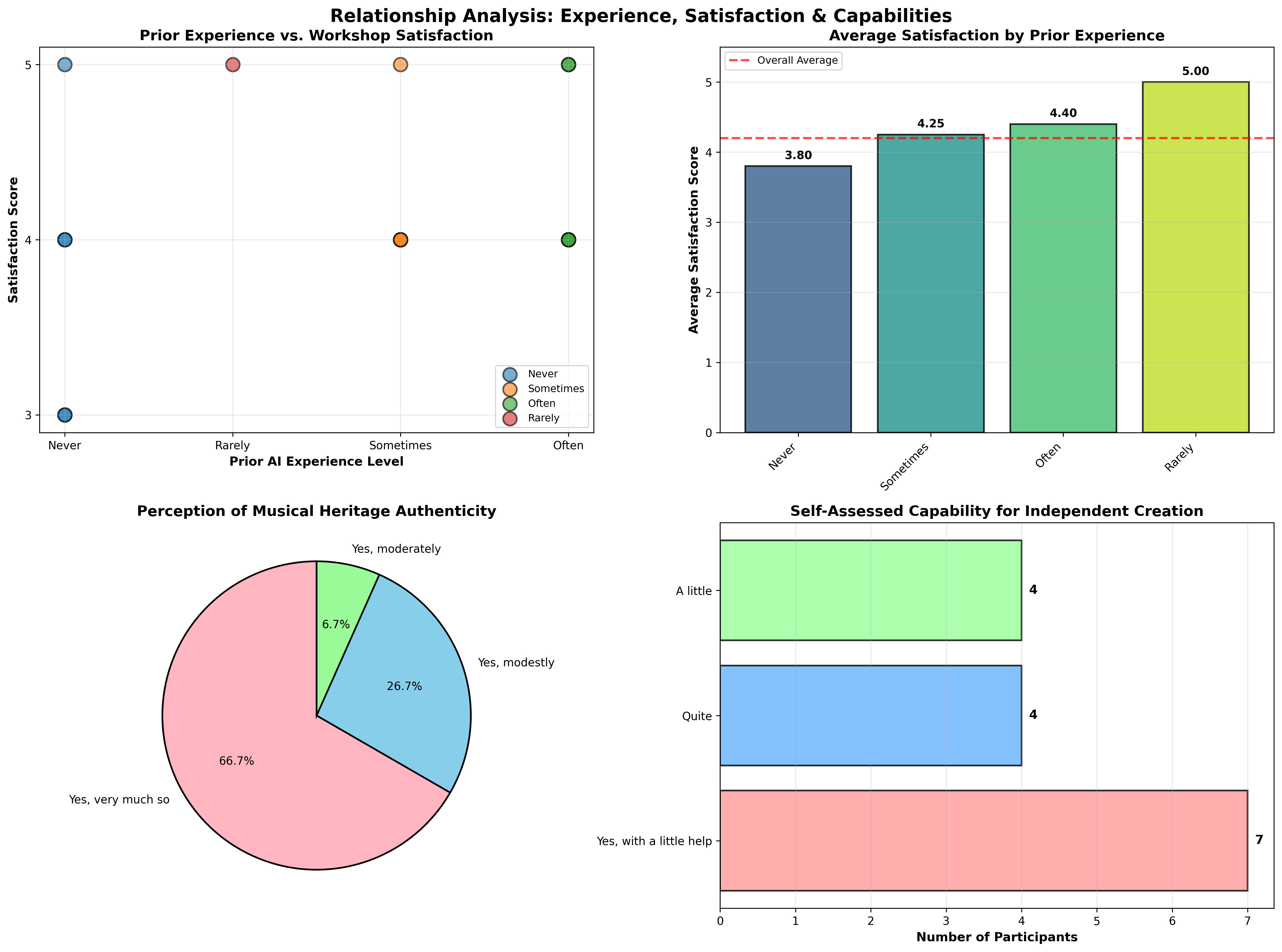}
  \caption{Relationship analysis showing prior AI experience vs. workshop satisfaction, average satisfaction by experience level, perception of musical heritage authenticity, and self-assessed capability for independent creation (n=13).}
  \label{relationship}
\end{figure}

\begin{figure}[h]
  \centering
  \includegraphics[width=0.8\textwidth]{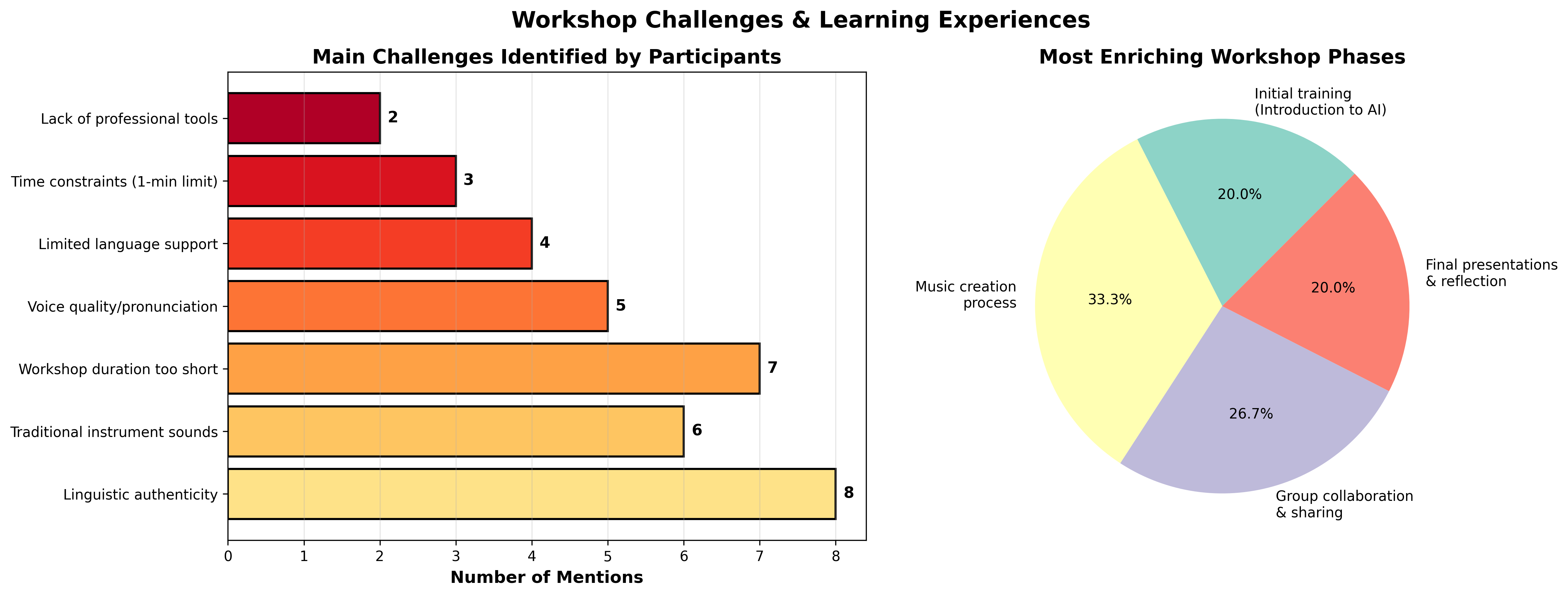}
  \caption{Workshop challenges identified by participants (left) and most enriching workshop phases (right). Linguistic authenticity and traditional instrument reproduction emerged as primary technical limitations.}
  \label{challenges}
\end{figure}

\subsection{Participant Survey Analysis}

To complement the qualitative observations, a post-workshop survey was conducted among 13 participants to quantify their satisfaction and perceptions of the creative process. Given the small sample size (n=13), these quantitative findings should be interpreted as exploratory indicators rather than statistically generalizable results. Participant survey results are illustrated in Figure~\ref{relationship} and Figure~\ref{challenges}.

The results revealed a high satisfaction rate (average score: 4.15/5), with 77\% having previously used AI tools and 100\% reporting a positive change in perception of generative AI applied to music. Approximately 85\% felt that the compositions authentically reflected Malian musical heritage, and 80\% stated they felt capable of creating new pieces autonomously or with minimal assistance after the training.

Survey Limitations: Several methodological limitations should be acknowledged: (1) the response rate of 43\% (13/30) may introduce self-selection bias; (2) participants who remained for the survey may have been those most engaged with the process; (3) the survey was administered immediately post-workshop, potentially capturing enthusiasm rather than sustained impact; (4) social desirability bias may have influenced responses given the institutional nature of participants.

Qualitative comments emphasized the fusion of tradition and modernity as the most enriching part of the experience, and many noted that the use of national languages reinforced their cultural pride. However, participants also identified limitations, such as short workshop duration, imperfect pronunciation in local languages, and the need for more realistic instrument reproduction.

\section{Discussion}
\subsection{Transformative potential of Gen AI}

The results provided preliminary evidence that Gen AI, used in a culturally respectful participatory framework, may serve to catalyze the construction of peaceful narratives by linking ancestral memory and shared futures.

This approach is in line with contemporary reflections on cultural sovereignty and the role of technologies in preserving national creative identities. ``AI must be a lever of cultural sovereignty, putting technology at the service of creators and not the other way around''.~\cite{MinistereCulture2025}  Applied to the Malian context, this perspective positions Gen AI as a tool for amplifying local heritage and voices, contributing to a symbolic sovereignty where technology values endogenous cultures rather than standardizing them.

Alternative Explanations: The positive outcomes observed may be attributable to factors beyond the AI tools themselves. The workshop format brought together diverse stakeholders in a structured collaborative environment, which may independently foster dialogue and social cohesion regardless of the technological medium. The novelty effect of interacting with AI systems may also temporarily elevate enthusiasm. Additionally, the institutional visibility of the project (culminating in SENARE 2025 promotion) may have reinforced positive perceptions. Future research should employ control conditions to isolate the specific contribution of AI-assisted creation versus collaborative music-making more broadly.

\subsection{Limits Identified}

In the course of the workshop, participants encountered certain classes of limitations.

\textbf{Insufficient linguistic corpora:} some national languages lack sufficient corpora in technological tools, posing problems of transcription and tonal recognition.

\textbf{Algorithmic censorship:} certain words (e.g., \textit{Niger}, \textit{soko}) are automatically flagged or suppressed by moderation algorithms due to negative associations in other linguistic contexts.

\textbf{Authenticity-innovation tension:} the challenge of preserving traditional sound while innovating structurally, without falling into superficial hybridization.

\textbf{Ethical and legal issues:} participants expressed concerns about the origin of training data, intellectual property, and the remuneration of holders of traditional musical heritage.

\subsection{Implications for public policy}

Visibility of the workshop results on a national stage engendered a broader discussion with national authorities on the implications of the experiment.  Potential translational implications for public policy included:

\textbf{Institutional integration:} incorporation of Gen AI into national cultural programs with a clear ethical and legal framework.

\textbf{Educational curricula:} integration into music schools and cultural institutions to develop technological and creative skills.

\textbf{Linguistic infrastructure:} support for the construction of linguistic and musical corpora specific to national languages.

\section{Conclusion}
This exploratory study suggests the potential of generative artificial intelligence, properly framed and culturally contextualized, to contribute to heritage revitalization and the construction of peace narratives. Analysis of the twelve compositions produced reveals notable sophistication in musical hybridization, layered multilingualism, and progressive narrative construction.

Beyond musical production, even on the scale of a limited experiment, we have observed multidimensional transformative effects: identity strengthening, intergenerational and interinstitutional dialogue, and technological empowerment ensuring the replicability of the process. The institutional recognition at SENARE 2025 testifies to the social and political relevance of this approach.

This research is part of the reflections on digital cultural sovereignty, positioning Gen AI as a medium for promoting endogenous cultures. The initiative opens up future research perspectives: longitudinal studies assessing sustainable impact, comparative research with other post-conflict contexts, exploration of integration with other forms of cultural expression.
\bibliographystyle{splncs04}
\bibliography{references_refined}
\end{document}